\documentclass[twocolumn,showpacs]{revtex4}
\usepackage{graphicx}
\usepackage{times}
\newcommand{\ds}{\displaystyle}

\begin{document}

\title{Enhanced carrier scattering rates in dilute magnetic semiconductors with correlated impurities}

\author{F. V. Kyrychenko}
\author{C. A. Ullrich}
\affiliation{Department of Physics and Astronomy, University of Missouri, Columbia, Missouri 65211}

\date{\today}

\begin{abstract}
In III-V dilute magnetic semiconductors (DMSs) such as Ga$_{1-x}$Mn$_x$As, the impurity positions tend to be correlated, which
can drastically affect the electronic transport properties of these materials. Within the memory function formalism we have
derived a general expression for the current relaxation kernel in spin and charge disordered media and have calculated spin and
charge scattering rates in the weak-disorder limit. Using a simple model for magnetic impurity clustering, we find a significant
enhancement of the charge scattering. The enhancement is sensitive to cluster parameters and may be controllable through
post-growth annealing.
\end{abstract}


\maketitle
The perspective of utilizing charge {\em and} spin of the electrons
for new electronic device applications has generated tremendous interest in the field of spintronics
\cite{spintronic}. A unique combination of magnetic and semiconducting properties makes dilute magnetic
semiconductors (DMSs) very attractive for various spintronics applications \cite{ohno}. Among the family of DMSs,
much attention has been paid to Ga$_{1-x}$Mn$_x$As
since the discovery of its relatively high ferromagnetic transition temperature \cite{ohno}, with a current
record of $T_c = 159$ K \cite{record}.

In Ga$_{1-x}$Mn$_x$As, unlike in II-VI DMSs, the magnetic ions in substitutional positions act as acceptors delivering
one hole per ion. All Ga$_{1-x}$Mn$_x$As samples are, however, heavily compensated, with hole concentrations much
less than $x$. This signals the presence of substantial amounts of donor defects like arsenic antisites $\rm As_{Ga}$ or
interstitial manganese ions $\rm Mn_I$ generated during low temperature molecular beam epitaxial growth \cite{LT}. The magnetic and
transport properties of Ga$_{1-x}$Mn$_x$As depend not only on the manganese fraction $x$ but are extremely sensitive
to detailed growth conditions \cite{shimizu}, as well as to temperature and speed of post-growth annealing
\cite{hayashi,potashnik,yu}. This sensitivity points to the crucial role played by the defects and their
configuration, and has stimulated intense research on the structure of defects and their influence
on the magnetic and transport properties of DMSs \cite{timm,cui}.

Most theoretical models for transport in DMSs assume random defect distributions. However, the presence of both positively and
negatively charged defects results in a correlation of their positions. Indeed, Timm {\em et al.} \cite{timm} found in the limit
of thermal equilibrium that, driven by Coulomb attraction, the defects tend to form clusters. The main effect of clustering is
{\em ionic} screening of the disorder Coulomb potential, which has been shown to be necessary to correctly reproduce the band
gap, metal-insulator transition and shape of the magnetization curve \cite{timm}.

In this paper, we demonstrate that the correlation of defect positions can have a dramatic effect on electronic transport in
DMSs: the conductivity of Ga$_{1-x}$Mn$_x$As is strongly modified through a momentum dependent impurity structure factor. We will
show that the clustering significantly increases the charge scattering relaxation rate. At the same time, positional correlation
taken alone is  not sufficient to affect spin scattering: orientational correlation of spin scatterers is also necessary.

To describe the transport properties of DMSs we have employed the memory function formalism \cite{Gotze81,belitz,UV}. The central
point of this approach is the calculation of the current relaxation kernel (or memory function), whose imaginary part can be
associated with the Drude relaxation rate. To get an expression for the memory function in spin- and charge-disordered media we
have used the equation of motion approach \cite{Gotze72,GV} (details of the derivation will be presented elsewhere). Here, we are
particularly interested in a paramagnetic system in the long-wavelength limit, the case relevant for studying the conductivity in
DMSs above $T_c$. In this case the memory function is obtained as
\begin{eqnarray}\label{finalS}
M(\omega)  & = & \frac{V^2}{n m \omega}\sum_{ \bf k \atop \mu\nu } k_{\alpha} k_{\beta}
  \left\langle\hat{\cal{U}}_{\mu}({\bf -k})\;
  \hat{\cal{U}}_{\nu}({\bf k})\right\rangle_{H_m} \nonumber \\
  && \times
  \Big(\chi_{\rho^{\mu}\rho^{\nu}}({\bf k},\omega)-\chi^c_{\rho^{\mu} \rho^{\nu}}({\bf k},0) \Big),
\end{eqnarray}
where $n$ is the carrier concentration and the operators $\hat{\rho}^{\mu}$ are defined through a four-com\-ponent
($\mu=1,+,-,z$) charge and spin density vector,
\begin{equation}
  \hat{\rho}^{\mu}({\bf k})=\frac1{V}\sum_{\bf q} \sum_{\tau \tau'} (\sigma^{\mu})_{\tau \tau'} \;
  \hat{a}^+_{{\bf q-k},\tau} \, \hat{a}_{{\bf q},\tau'} \:.
\end{equation}
Here, $\sigma^\mu$ is defined via the Pauli matrices, where $\sigma^1$ is the $2\times 2$ unit matrix, $\sigma^\pm=(\sigma^x \pm
i \sigma^y)/2$, and $\chi_{\rho^{\mu} \rho^{\nu}}({\bf k},\omega)$ are the associated charge- and spin-density response
functions. The superscript $c$ in Eq. (\ref{finalS}) refers to a clean (defect-free)  system.

The presence of impurities, including their correlations, enters in Eq. (\ref{finalS}) through the expression
$\langle\hat{\cal{U}}_{\mu}({\bf -k})\hat{\cal{U}}_{\nu}({\bf k})\rangle_{H_m}$. The angular brackets indicate a thermodynamical
average with respect to a magnetic subsystem Hamiltonian $\hat{H}_m$. We assume $\hat{H}_m$ to be a sum of individual spin
contributions corresponding to uncorrelated and noninteracting localized spins. The disorder scattering potential is described by
the four-component impurity charge- and spin-density operator
\begin{equation}\label{pot}
  \hat{\cal{U}}({\bf k})=\frac1{V}\sum_j \left(\begin{array}{c}
    U_1({\bf k}) \\
    \frac{J}2 \hat{S}^j_- \\
    \frac{J}2 \hat{S}^j_+ \\
    \frac{J}2 \hat{S}^j_z \
  \end{array} \right) e^{i{\bf k\cdot R}_j},
\end{equation}
where the summation is performed over all defect positions. In order to separate the effect
of the impurity structure factor from other effects of clustering like ionic screening we
consider only one type of defects, $\rm Mn^{2+}$ ions in cation substitutional positions.
All defects in Eq.~(\ref{pot}) produce thus the same charge potential and carry localized
spins. The latter are treated as quantum mechanical operators coupled to the band
carriers via a contact Heisenberg interaction resulting in a momentum-independent
exchange constant $J$. In our calculations we neglect the time evolution of localized spin
operators, thus assuming that the carriers move through an ensemble of frozen spins. For the charge
component $U_1({\bf k})$ we take a Coulomb potential screened with the host material
dielectric constant. Effects of ionic screening are disregarded, and screening by the
electron liquid is absorbed in the response functions.

Let us now consider the correlated product of two components of the disorder potential in Eq.~(\ref{finalS}). First, we separate
the same-ion ($j=j'$) and pair ($j\neq j'$) contributions:
\begin{eqnarray}\label{same}
&&\left\langle\hat{\cal{U}}_{\mu}({\bf -k})\; \hat{\cal{U}}_{\nu}({\bf k})\right\rangle_{H_m}=
  \frac{n_i}V \left\langle\hat{U}_{\mu}({\bf -k})\; \hat{U}_{\nu}({\bf k})\right\rangle_{H_m} \\
  &+& \frac1{V^2} \left\langle\hat{U}_{\mu}({\bf -k})\right\rangle_{H_m} \left\langle\hat{U}_{\nu}({\bf k})\right\rangle_{H_m}
  \sum_{j \neq j'} \;e^{i{\bf k}\cdot({\bf R}_{j'}-{\bf R}_{j})}, \nonumber
\end{eqnarray}
where $n_i$ is the impurity concentration, and in the second term the average of the product becomes a product of averages for
noninteracting spins. Random impurities are taken into account through the first term on the right-hand side of Eq.~(\ref{same}),
while the second term contributes only if there are correlations in the impurity positions. For spin scattering, however, spatial
correlations are not sufficient for the pair term to survive. Indeed, if any of the indices $\mu,\nu$ corresponds to a spin
component (e.g., $\mu=z$), then the second term in Eq.~(\ref{same}) is proportional to the average spin and vanishes if $\langle
\hat{S}_z \rangle=0$, regardless of spatial correlations. The presence of {\em two} sources of randomness is a characteristic
feature of spin scattering. In terms of scattering, localized spins are correlated if they exhibit {\em both} positional and
orientational correlations. Note that the presence of a macroscopic magnetization is not necessary for spins to be correlated.
What counts in the scattering is the short-range orientational correlation that might be present even in a macroscopically
paramagnetic system.

For the charge-scattering term ($\mu,\nu=1$), we have
\begin{equation}
  \left\langle\hat{\cal{U}}_1({\bf -k})\; \hat{\cal{U}}_1({\bf k})\right\rangle_{H_m}=
  \left|U_1({\bf k})\right|^2 \frac{n_i}V \: S({\bf k}),
\end{equation}
with the structure factor
\begin{equation}\label{factor}
  S({\bf k})=1+\frac{\Omega_0}{V x} \sum_{j \neq j'} \;e^{i{\bf k}\cdot({\bf R}_{j'}-{\bf R}_{j})},
\end{equation}
where $x$ is the molar fraction of magnetic ions in the sample and $\Omega_0$ is the elementary cell volume. Let us introduce a pair
distribution function $P({\bf R})$,  normalized as
\begin{equation}\label{norm}
  \frac{1}V \int_V P({\bf R}) d{\bf R} = x,
\end{equation}
which describes the probability to find another magnetic ion at a distance ${\bf R}$ from a given ion.
We approximate Eq.~(\ref{factor}) as
\begin{equation}\label{factor1}
  S({\bf k})\approx 1+\frac1{\Omega_0} \int_V P({\bf R}) \cos\left({\bf k\cdot R}\right) d{\bf R}.
\end{equation}
For a random impurity distribution one has $P({\bf R})=x$, and the second term in Eq.~(\ref{factor1}) vanishes.
The structure factor $S({\bf k})$ is then equal to 1, which implies a contribution only of the same-ion term in Eq.~(\ref{same}).

To study the effect of correlations in the defect positions (higher probability to find magnetic impurities close to each
other), we employ a simple model expression for a pair distribution function $P({\bf R})$, assuming it to be a piecewise constant,
spherically symmetrical function of the form
\begin{equation}\label{P}
  P(R)=\left\{\begin{array}{ll}
     \ds x_c, & R<R_c, \\
     \ds x_d, \hspace{0.7cm} & R_c<R<R_d  \\
    x, & R>R_d \:. \
  \end{array} \right.
\end{equation}
The first region corresponds to a cluster of radius $R_c$ with effective impurity
concentration $x_c>x$, the second region is a depletion layer with $x_c<x$,
necessary to preserve the average impurity concentration in the sample. The width of the
depletion layer is determined by the normalization condition (\ref{norm}). The impurity
concentration within the depletion layer has a minor impact on the final results, and to
keep things simple we fix it to $x_d=x/2$.

The remaining two parameters, $R_c$ and $x_c$, describe the cluster structure and are in general independent. We can relate them,
however, if we fix the average number $N$ of the impurity ions within the cluster:
\begin{equation} \label{relate}
\frac{4 \pi R_c^3}{3 \Omega_0}x_c=N.
\end{equation}
This seems reasonable for modelling the effect of annealing on low-temperature grown DMS samples, where one may
assume that the total number of $\rm Mn^{2+}$ substitutional ions within the cluster is conserved while the
cluster size (and density) may vary. In the following we will use $N=10$, consistent with the results of Monte-Carlo
calculations \cite{timm}.

\begin{figure}
\begin{center}\leavevmode
\includegraphics[width=0.9\linewidth]{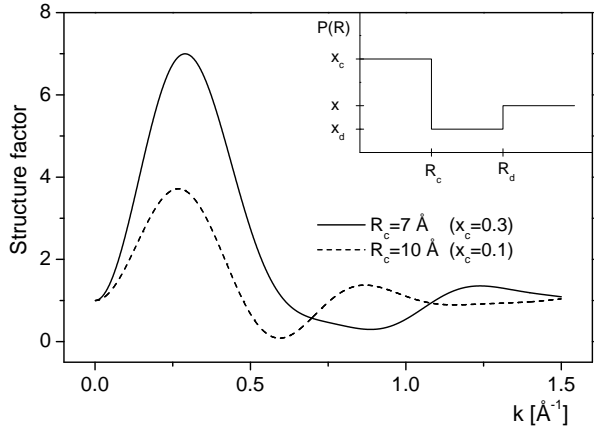}
\caption{Momentum-dependent impurity structure factor (\ref{Sk}) for different values of cluster radius $R_c$. Number of ions
within the cluster $N=10$, average impurity concentration in the sample $x=0.05$. Inset: schematic plot of the pair distribution
function (\ref{P}). }\label{structure}
\end{center}
\end{figure}

Our model pair correlation function (\ref{P}) yields the following
momentum-dependent impurity structure factor:
\begin{eqnarray}
  S(k) &=& 1+\frac{2 \pi x}{k^3\Omega_0}\left\{\frac{2x_c - x}{x}[\sin(k R_c)
  - k R_c \cos(k R_c)]\right. \nonumber \\
  &&
   -\left[\sin(k R_d) - k R_d \cos(k R_d)\right]\bigg\}.  \label{Sk}
\end{eqnarray}
Fig.~\ref{structure} shows $S(k)$ for two different values of the cluster radius $R_c$, which
in principle can be controlled by annealing. As expected, the structure factor oscillates with decreasing amplitude,
with a larger first maximum for smaller cluster size.

Eq.~(\ref{finalS}) contains the set of full system charge- and spin-density response functions and, strictly speaking, should be
calculated by iterations. This approach was realized in \cite{Gold} to study a spin-independent system close to the metal
insulator threshold. In our case, however, we assume that the disorder is weak enough so we can approximate Eq.~(\ref{finalS}) by
expanding to second order in the disorder potential $\hat{\cal{U}}({\bf k})$, and thus replace $\chi_{\rho^{\mu}\rho^{\nu}}({\bf
k},\omega)$ by its clean system counterpart $\chi^c_{\rho^{\mu}\rho^{\nu}}({\bf k},\omega)$.

An accurate description of carrier-mediated ferromagne\-tism \cite{dietl} and optical response \cite{texas} of Ga$_{1-x}$Mn$_x$As
would require taking the true multiband structure of the material into account. In this paper, however, we concentrate primarily
on effects arising from the disorder configuration, putting less emphasis on the details of the band structure.
In the evaluation of the response functions $\chi^c_{\rho^{\mu}\rho^{\nu}}({\bf k},\omega)$ we thus work with a simple parabolic
band.

Furthermore, we use the random phase approximation (RPA) to account for
electron-electron interactions. For simplicity we limit ourselves
to a static RPA and express the density-density response function of the interacting system as
\begin{equation}
  \chi_{nn}({\bf q},\omega) = \frac{\chi_0({\bf q},\omega)}{\varepsilon_{\rm RPA}({\bf q},0)} \:,
\end{equation}
where $\chi_0({\bf q},\omega)$ is the non-interacting density-density response function of the clean system, i.e., the Lindhard
function, and $\varepsilon_{\rm RPA}({\bf q},0)$ is the static RPA dielectric function \cite{mahan}. The spin response functions
in the paramagnetic state are not affected by electron-electron interactions on the RPA level, and they can also be expressed in
terms of the Lindhard function $\chi_0({\bf q},\omega)$.

Under the above approximations Eq.~(\ref{finalS}) can be directly evaluated and the memory function becomes the sum of spin and
charge contributions
\begin{equation}
  M(\omega)=\frac1{\tau_n}+\frac1{\tau_s},
\end{equation}
with
\begin{equation}\label{taun}
  \frac1{\tau_n(\omega)}= A \int\limits_0^{\infty}\!  k^4 S(k)
  \frac{|U_1(k)|^2}{\varepsilon_{\rm RPA}(k)}  \frac{\chi_0(k,\omega)-\chi_0(k)}{\omega} \, dk,
\end{equation}
and
\begin{equation}\label{taus}
  \frac1{\tau_s(\omega)} = A \, \frac{J^2}{4}\, S_{\rm Mn}(S_{\rm Mn}+1)\int\limits_0^{\infty} k^4 \,
  \frac{\chi_0(k,\omega)-\chi_0(k)}{\omega} \, dk.
\end{equation}
Here $S_{\rm Mn}=5/2$ is the localized spin of magnetic impurities, and the common prefactor is given by $A=(n_i/n)V^2/6\pi^2 m$.
The imaginary parts of Eqs.~(\ref{taun}) and (\ref{taus}) represent the energy dependent charge- and spin-scattering
contributions to the Drude relaxation rate. Generally speaking, within our model the relaxation times $\tau_{n(s)}(\omega,{\bf
q})$ are also momentum dependent. In the present paper, however, we consider only a long-wavelength limit, setting ${\bf q}\to
0$. Note that the momentum-dependent impurity structure factor $S(k)$ does not appear in the spin term  (\ref{taus}).

We have evaluated the scattering rates, Eqs.~(\ref{taun}) and (\ref{taus}), for the case of $\rm Ga_{0.95}Mn_{0.05} As$ with a
hole concentration of $p=0.5$ hole per magnetic ion. Other parameters used are: heavy hole effective mass $m=0.5\, m_0$,
dielectric constant $\varepsilon=13$, and exchange constant $VJ=55\:{\rm meV\,nm^3}$, which corresponds to the widely used DMS
p-d exchange constant $N_0 \beta=1.2\,$eV \cite{dietl}.

\begin{figure}
\begin{center}\leavevmode
\includegraphics[width=1.0\linewidth]{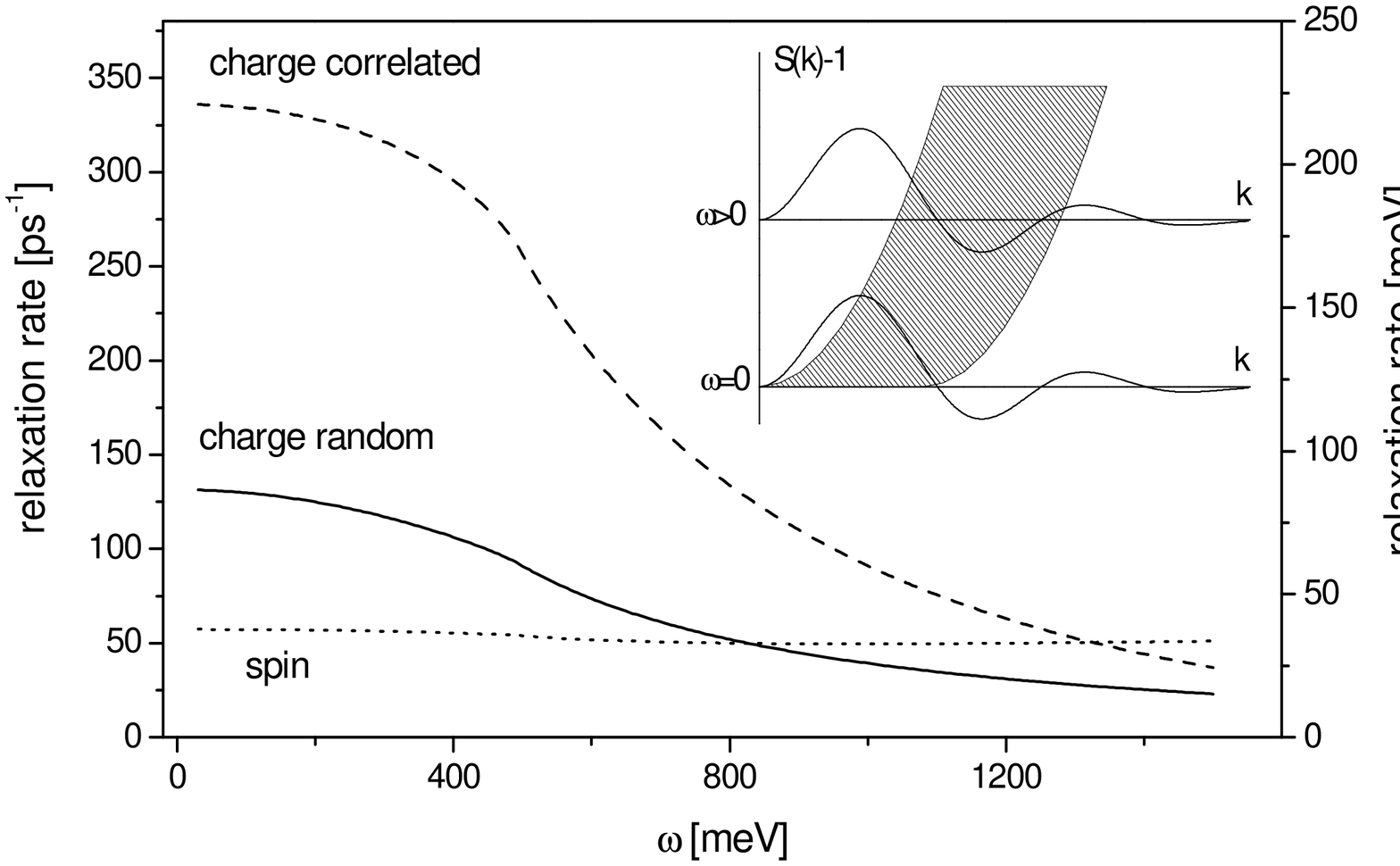}
\caption{Frequency dependence of charge (\ref{taun}) and spin (\ref{taus}) scattering relaxation rates, for impurity clusters
with $R_c=10$\AA, $x_c=0.1$. Inset: illustration of the frequency dependence of the
cluster enhancement, see text.} \label{relax}
\end{center}
\end{figure}

In Fig.~\ref{relax} we plot the frequency dependence of the imaginary part of Eqs.~(\ref{taun}) and (\ref{taus}). Both spin and
charge relaxation rates demonstrate the frequency dependencies that one might expect for momentum-independent scattering and for
Coulomb scattering. In agreement with earlier estimations by golden rule \cite{golden}, the charge scattering in dc limit
dominates the spin scattering. We would like to point out, however, that in our calculations (as well as in Ref.~\cite{golden})
we did not take into account the effect of ionic screening shown \cite{timm} to reduce significantly the charge disorder
potential. In any case, even in the present model the spin scattering contribution is clearly not negligible, and reaches the
same order of magnitude as charge scattering for higher frequencies. Both contributions should therefore be taken into account
simultaneously.

Correlation in impurity positions results in a significant increase of the charge scattering contribution at low frequency, while for
higher $\omega$ this enhancement decreases. The origin of this effect is illustrated in the inset in Fig.~\ref{relax}. The
presence of impurity clusters selects excitations within a finite momentum window defined by the impurity structure factor.
On the other hand, the region of one-particle excitations is given by the imaginary part of the Lindhard function $\chi_0$
(shaded region in Fig.~\ref{relax}). The maximum enhancement of the relaxation rate corresponds to maximum overlap of the
selection window with the one-particle spectrum, which happens at low frequency. For higher $\omega$ the window falls outside of
the available excitation spectrum, reducing the relaxation rate enhancement.

\begin{figure}
\begin{center}\leavevmode
\includegraphics[width=1.0\linewidth]{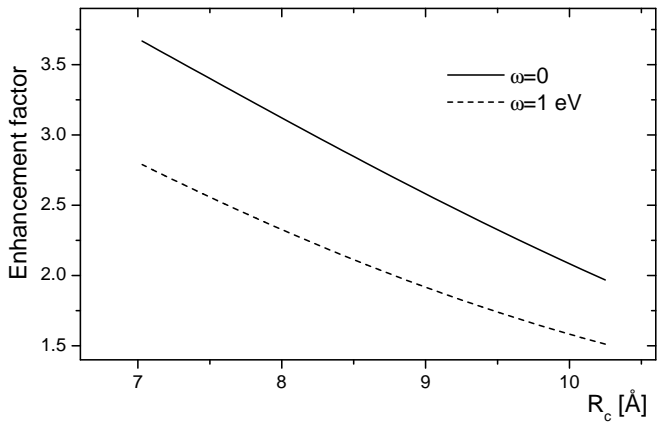}
\caption{Relaxation rate enhancement (\ref{xi}) due to correlation in impurity positions  as a function of cluster size.
The average number of magnetic ions within the cluster is fixed to $N=10$.}\label{enhanc}
\end{center}
\end{figure}

In Fig.~\ref{enhanc} we plot the cluster enhancement factor
\begin{equation}\label{xi}
  \xi=\frac{\tau_n^{-1}+\tau_s^{-1}}{(\tau_n^R)^{-1}+(\tau_s^R)^{-1}}
\end{equation}
as a function of cluster configuration. The enhancement factor is defined as the ratio of the total (charge plus spin) relaxation
rates for correlated and random impurity distributions. Recall that the average number of magnetic ions within
the cluster is fixed to $N=10$, and the cluster radius $R_c$ is related to concentration of magnetic ions within the cluster
$x_c$ through Eq.~(\ref{relate}). The enhancement is strongest (up to a factor of 3) for low frequency, and is quite sensitive to
cluster configuration. The latter will be sensitive to post-growth annealing, which is widely used to increase $T_c$ in
Ga$_{1-x}$Mn$_x$As samples. The possible modification of transport properties described here should therefore be taken into
account along with other effects of annealing like decrease of the number of interstitial Mn ions and increase of carrier
concentration.

To summarize, using the memory function formalism we have considered transport in charge and spin disordered media with potential
application for DMSs, with particular emphasis on non-randomness of impurity positions in Ga$_{1-x}$Mn$_x$As. We have shown that
positional correlations alone of the magnetic impurities do not affect spin scattering: orientational correlations are also
necessary. For charge scattering, impurity clustering gives rise to a momentum-dependent impurity structure factor which
substantially modifies the transport properties of the material, typically leading to 100\% enhancements of relaxation rates.
These results should give valuable insights into the effects  of annealing on low temperature grown DMS samples.

Finally, the discussion in this paper was limited to DMS in the paramagnetic state. However, it is well known that
magnetic ordering in DMS can have a dramatic influence on transport properties \cite{omiya}. Our approach
should be well suited to study these effects.

\acknowledgments
This work was supported by DOE Grant No. DE-FG02-05ER46213.

\end{document}